\documentstyle[11pt,epsf,psfig]{article}

\newcommand{\del}{\partial}
\newcommand{\beq}{\begin{eqnarray}}
\newcommand{\eeq}{\end{eqnarray}}
\newcommand{\be}{\begin{eqnarray*}}
\newcommand{\ee}{\end{eqnarray*}}

\newcommand{\ra}{\rightarrow}

\newcommand{\ve}{\varepsilon}
\newcommand{\nn}{\nonumber}

\newcommand{\lam}{\bar{\lambda}}

\begin{document}

\centerline{\Large\bf {A more accurate analysis of}}
\medskip
\centerline{\Large\bf {Bose-Einstein condensation in harmonic traps}}
\vskip 10mm
\centerline{H. Haugerud, T. Haugset and F. Ravndal}
\centerline{\it Institute of Physics}
\centerline{\it University of Oslo}
\centerline{\it N-0316 Oslo, Norway}

\bigskip
{\bf Abstract:} {\small Using the Euler-Maclaurin summation we calculate analytically 
the internal energy for non-interacting bosons confined within a harmonic oscillator 
potential. The specific heat shows a sharp $\lambda$-like peak indicating a 
condensation into the ground state at a well-defined transition temperature. Full 
agreement is obtained with  direct numerical calculation of the same quantities.
When the number of trapped particles is very large and at temperatures near and above
the transition temperature, the results 
also agree with previous approximate calculations. At extremely low temperatures both the
specific heat and the number of particles excited from the condensate are exponentially
suppressed.}
  
\vspace{10mm}
Bose-Einstein condensation  has now been experimentally demonstrated in magnetic
traps of rubidium\cite{rub}, lithium\cite{lit} and most recently sodium\cite{sod} gases.
To a good approximation one describes the trapping potentials as 3-dimensional anisotropic
oscillators which in the rubidium experiments\cite{rub} have frequencies typically 
around $\omega = 500 - 1000$ Hz. For the sake of simplification we will here ignore the
anisotropy and the weak interactions between the alkali gas atoms. The energy levels
of one particle are then simply given as $\ve_n = \hbar\omega n$ where the quantum 
number $n$ takes the values $n = 0,1,2\ldots{}$ when we drop the zero-point energy.
Since each energy level has a degeneracy $g_n = (n+1)(n+2)/2$, the total number
$N$ of particles in such a trap at temperature $T$ and chemical potential $\mu$ is 
given by the Bose-Einstein distribution as
\beq
     N = \sum_{n=0}^\infty {g_n\over e^{\beta(\ve_n - \mu)} - 1}
\eeq
where $\beta = 1/k_BT$. The ground state has quantum number $n = 0$ and thus contains
$N_0 = \lambda/(1 - \lambda)$ particles where $\lambda = \exp{(\beta\mu)}$ is the 
fugacity. We then have
\beq
      N = {\lambda\over 1 - \lambda} + N_e                          \label{N}
\eeq
where the number of particles in the higher states is
\beq
     N_e = \sum_{n=1}^\infty {g_n\lambda e^{-bn}
              \over 1 - \lambda e^{-bn}}                         \label{sum1}
\eeq
It depends only on the effective fugacity $\lam = \lambda \exp{(-b)}$ where 
$b = \hbar\omega/k_BT$.

When the temperature is lowered, the fugacity $\lambda \ra 1$ and a finite fraction of 
particles starts to condense into the ground state.  Since the total number of particles 
$N \gg 1$, this will happen when the variable $b \ll 1$. In this region and
at higher temperatures one can then approximate the sum in 
(\ref{sum1}) by an integral. Including  only the leading term $\sim n^2$ in the
degeneracy, one then has the semiclassical limit\cite{HH}\cite{DK}.  
Defining the transition temperature to be the temperature at which the fugacity takes 
the value $\lambda = 1$, it is then found\cite{HH} to be given by 
\beq
     T_0 = {\hbar\omega\over k_B} \left({N\over \zeta(3)}\right)^{1/3}   \label{T0}
\eeq
The resulting specific heat exhibits a sharp jump with the shape of a $\lambda$. This is
in contrast with the gas in zero potential where the specific heat is continuous. As
a check, the partition function was also calculated  using numerical summation.

A more accurate approximation of the sum has been given by Grossmann and
Holthaus \cite{GH}. Based upon the two leading terms in the degeneracy, they constructed
a continuous density of states which then made it possible to approximate the sum by two
integrals. This caused a small shift in the transition temperature 
by a factor $\sim N^{-1/3}$ (\ref{T0}) and a corresponding small rounding-off of the 
sharp peak in the 
specific heat when the particle number $N$ is not too large. In the experiments under 
consideration, $N$ takes typically values in the range from  $10^3$ to $10^6$.

Recently, these approximation methods in which the sum is replaced by integrals,
have been criticized by Kirsten and Toms\cite{DT}. Instead, they propose to evaluate
the sum directly by contour integration. In turns out, however, that this method is 
difficult to use in the transition region where they are forced to fall back upon a 
direct, numerical summation. Their results are thus quite similar to those of Grossmann and
Holthaus\cite{GH}.

The standard method for summing a series like (\ref{sum1}) is  Euler-Maclaurin
summation\cite{NR}. One then has
\beq
    \sum_{n=a}^{n=b}f(n) = \int_{x=a}^{x=b}dxf(x) + {1\over 2}[f(b) + f(a)]
    + {1\over 12}[f'(b) - f'(a)]  + \cdots{}   \label{EM}
\eeq
In our case the contributions from the upper limit will vanish. When
$T \gg \hbar\omega/k_B$ we can also safely ignore the contributions coming from the higher
order derivatives at the lower integration limit. The sum (\ref{sum1}) can then be
written as
\beq
    N_e = G_3(\lam) + {3\over 2}G_2(\lam) + G_1(\lam)
 + {1\over 4}{\lam\over 1 - \lam}\left({31\over 6} + {b\over 1 - \lam}\right)  \label{sum2}
\eeq
when we only include terms up to the first derivative. We have here introduced the new
functions
\beq
    G_{p+1}(\lam) = {1\over p!}\int_1^\infty dx 
                     {x^p\lambda e^{-bx}\over  1 - \lambda e^{-bx}}           \label{Gs}
\eeq
In the case of Bose-Einstein condensation in zero potential, the lower limit would have
been $x = 0$ and the functions would be equal to the $g_p(\lambda)$ 
functions\cite{Pathria}. These are equal to the more standard polylogarithmic functions
\beq
     Li_p(z) = \sum_{n=1}^\infty {z^n\over n^p}
\eeq
We see that $Li_1(z)= - \ln{(1-z)}$. Furthermore we will need $Li_p(1) = \zeta(p)$.
By a partial integration  one finds that the functions (\ref{Gs}) can be obtained
from the simple recursion relation
\beq
        G_{p+1}(\lam) = {1\over bp!}Li_1(\lam) 
                + {1\over b}\int_0^{\lam} {d\lambda\over\lambda}G_p(\lambda)
\eeq
Since derivatives of polylogarithmic functions satisfy $Li^{\;'}_{p+1}(z)= Li_p(z)/z$, 
the new functions (\ref{Gs})  can all be expressed by these. 
With the result (\ref{sum2}) for the number of excited particles we then have
\beq
     N &=& {\lambda\over 1 - \lambda}  + {1\over b^3}Li_3(\lambda e^{-b}) 
       + {5\over 2b^2}Li_2(\lambda e^{-b}) + {3\over b}Li_1(\lambda e^{-b})   \nn    \\
      &+&  {1\over 4}{\lambda e^{-b}\over 1 - \lambda e^{-b}}
      \left({31\over 6} + {b\over 1 - \lambda e^{-b}} \right)           \label{sum3}
\eeq

This equation determines the critical behaviour of the system.

\begin{figure}[htb] 
\begin{center}
\mbox{\psfig{figure=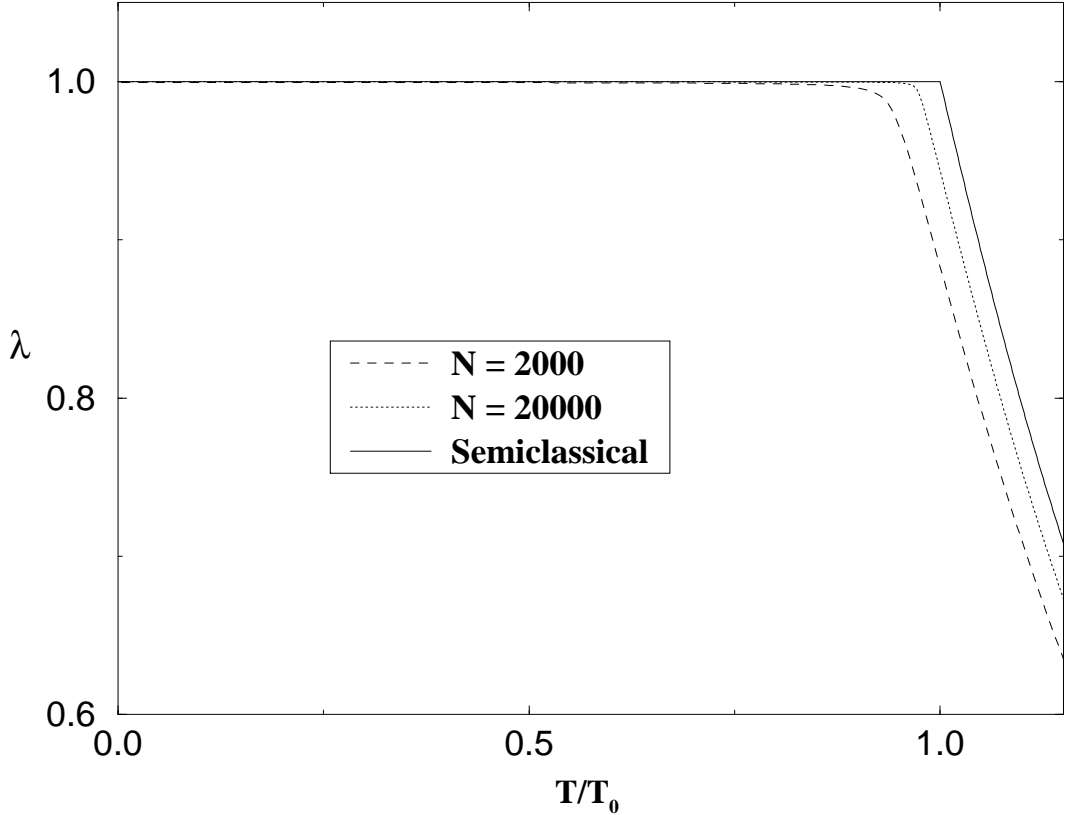,width=14cm,angle=270}}
\end{center}               
\caption[b]{\protect\footnotesize The fugacity $\lambda$ as a function of $T/T_0$ for $N = 2000$ and $N = 20000$. $\lambda$ is close to unity at low temperature and starts to fall off rapidly around $T=T_0$. As $N$ is increased, the transition sharpens and the curve approaches the semiclassical result. }
\label{LAM}                  
\end{figure}

\begin{figure}[htb] 
\begin{center}
\mbox{\psfig{figure=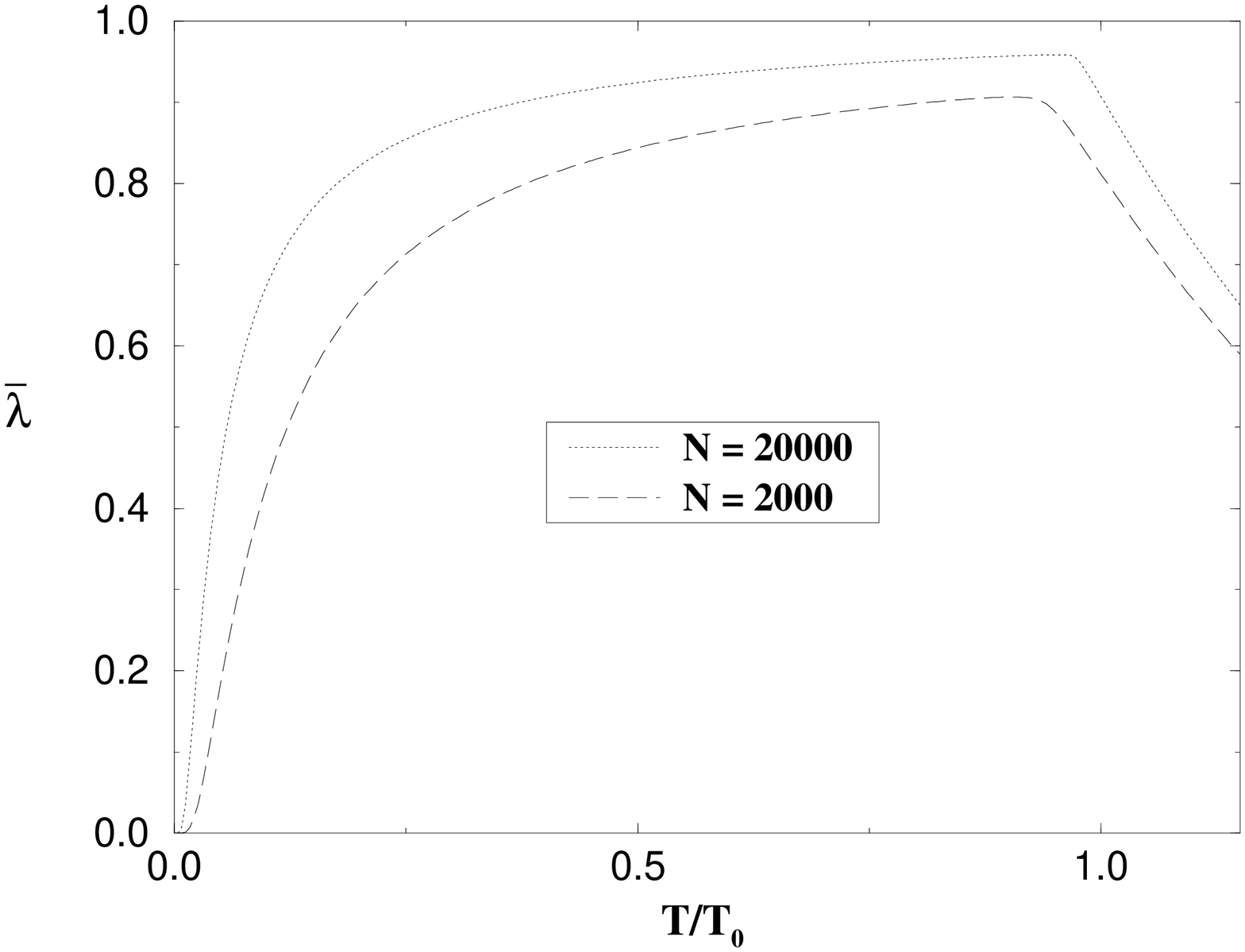,width=14cm,angle=270}}
\end{center}
\caption[b]{\protect\footnotesize The effective fugacity $\bar{\lambda}$ as a function of $T/T_0$ for $N = 2000$ and $N = 20000$. $\bar{\lambda}$ is always less than unity and vanishes at absolute zero. The maximum reached near the transition is seen to sharpen as $N$ is increased.}
\label{LAMEFF}                                        
\end{figure}

At temperatures below the semiclassical transition temperature (\ref{T0}) most of the particles
are in the condensate consisting of $N_0$ particles in the ground state of the harmonic
oscillator. The fugacity $\lambda \simeq 1$  follows from
\beq
     \lambda = {N_0\over N_0 + 1}                                              \label{L}
\eeq
when $N_0 \gg 1$. For higher temperatures we must use the full equation (\ref{sum3}) to 
calculate $\lambda$.
The result is shown in Fig. \ref{LAM} for $N = 2000$ and $N = 20\,000$ particles. We see that when the
number of particles gets to be very large, the transition into the ground state marked by having
$\lambda = 1$, becomes correspondingly sharp. In contrast, the effective 
fugacity $\lam < 1$ at all temperatures as is clearly seen from Fig. \ref{LAMEFF}.
The number of excited particles and as we will see, also the specific heat will thus be 
exponentially suppressed at very low temperatures.

Above the transition temperature (\ref{T0}) the two fugacities can be taken to be equal to a very 
good accuracy. This corresponds to having the parameter $b \ll 1$ in this temperature region. The
leading terms in the number of excited particles $N_e$ are the second and third terms in 
(\ref{sum3}), i.e.
\beq
     N_e = {1\over b^3}Li_3(\lambda e^{-b}) + {5\over 2b^2}Li_2(\lambda e^{-b})  + {\cal O}(1/b)
\eeq
Expanding the polylogarithms around $b=0$ we then get
\beq
     N_e = {1\over b^3}Li_3(\lambda) + {3\over 2b^2}Li_2(\lambda)  + {\cal O}(1/b)  \label{approx}
\eeq
The two first terms here are the same as obtained in the calculation of  Grossmann and 
Holthaus\cite{GH}. They were obtained from a density of states based upon only the first two terms 
in the quantum mechanical degeneracy $g_n = (n^2 + 3n + 1)/2$. Their results are in surprisingly
good agreement with the exact numerical results near and above the transition temperature. This is
to a large extent explained by an apparent compensation of their neglect of the third term in the
degeneracy by taking the integration from $n=0$ instead of from $n=1$. Their results thus depend 
on the 
ordinary fugacity $\lambda$ instead of the effective fugacity $\lam$ which appears in our more accurate
analysis. As we have seen, it is the exponentially damped effective fugacity which determines
the thermodynamics at very low temperatures. 

In the transition region we can use the approximate result (\ref{approx}) for the number of
excited particles. The number of particles in the condensate thus varies with the temperature as
\beq
    N_0 = N - \left({k_BT\over\hbar\omega}\right)^3\zeta{(3)} 
        - {3\over 2}\left({k_BT\over\hbar\omega}\right)^2\zeta{(2)}              \label{cond}
\eeq
Defining the transition temperature $T_c$ where  $N_0 = 0$ we see that is is given by
\be
   T_c &=& T_0\left[ 1 
       - {3\zeta{(2)}\over 2N}\left({k_BT_C\over\hbar\omega}\right)^2\right]^{1/3} \nn \\
       &\simeq& T_0\left[1 - {\zeta{(2)}\over 2\zeta{(3)}^{2/3}}{1\over N^{1/3}}\right] \label{TC} 
\ee
as obtained by Grossmann and Holthaus\cite{GH}. When the number of particles in the trap 
becomes very large, the transition temperature approaches the semiclassical result
(\ref{T0}). Similarly, we get for the variation of the condensate (\ref{cond}) near the
transition temperature,
\beq
    {N_0\over N} = 1 -  \left({T\over T_0}\right)^3 
  - {3\zeta{(2)}\over 2\zeta{(3)}^{2/3}}{1\over N^{1/3}}\left({T\over T_0}\right)^2 
 \eeq
This result is only approximately correct near the transition region where the effective
fugacity $\lam \simeq 1$. A more accurate result is obtained by numerically solving (\ref{sum3}) for $\lambda$. 

\begin{figure}[htb] 
\begin{center}
\mbox{\psfig{figure=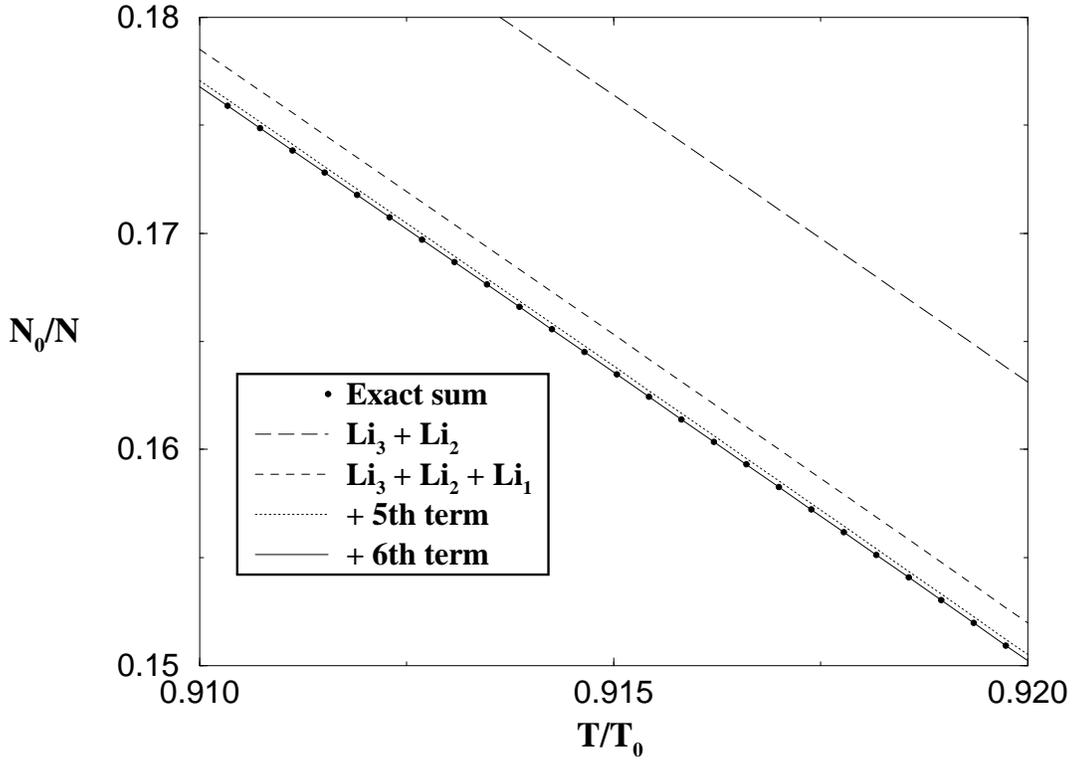,width=14cm,angle=270}}
\end{center}
\caption[b]{\protect\footnotesize The condensate ratio $N_0/N$ as a function of $T/T_0$ for $N=20000$ just below the transition. The approximate results are seen to approach the exact numerical results as more terms in Eq. (\ref{sum3}) are included. When the third derivative in the Euler-Maclaurin formula is included, a numerical agreement to five decimal places is obtained. }
\label{CONDENSATE1}                                        
\end{figure}

\begin{figure}[htb] 
\begin{center}
\mbox{\psfig{figure=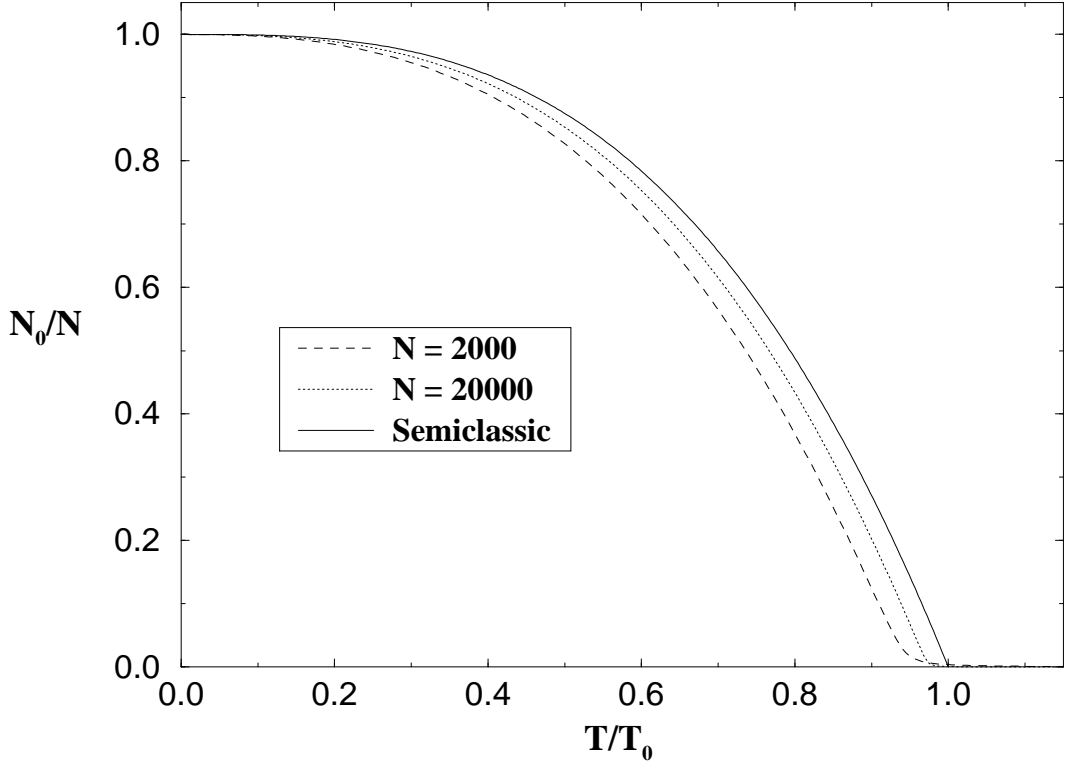,width=14cm,angle=270}}
\end{center}
\caption[b]{\protect\footnotesize The condensate ratio $N_0/N$ as a 
function of $T/T_0$ for $N = 2000$ and $N = 20000$. $N_0$ approaches $N$ exponentially at extremely low temperatures. Near $T=T_0$ the sharp transition found with the semiclassical approximation is rounded off by higher order corrections. In the figure one can not distinguish between the analytical and numerical results.}
\label{CONDENSATE2}                                        
\end{figure}

The result is shown in Fig. \ref{CONDENSATE1} which gives the variation of the condensate just below the transition temperature. We see there the limited accuracy of just including the first two terms in $N_e$ when comparing to the exact result obtained by numerical summation. Using the Euler-Maclaurin formula up to and including the third-derivative term, we find a numerical agreement to five decimal digits. In Fig. \ref{CONDENSATE2} we show the condensate varying with the temperature
down to absolute zero. As $T \ra 0$ the exponential suppression of excited particles
becomes stronger and stronger. In fact, when $T < \hbar\omega/k$,
essentially all the excited particles will be in the $n=1$ energy level of the oscillator. We then
have 
\beq
     N_0 = N - 3 e^{-\hbar\omega/k_BT}                                         \label{Ne}
\eeq
when we neglect more exponentially suppressed terms.

\begin{figure}[htb] 
\begin{center}
\mbox{\psfig{figure=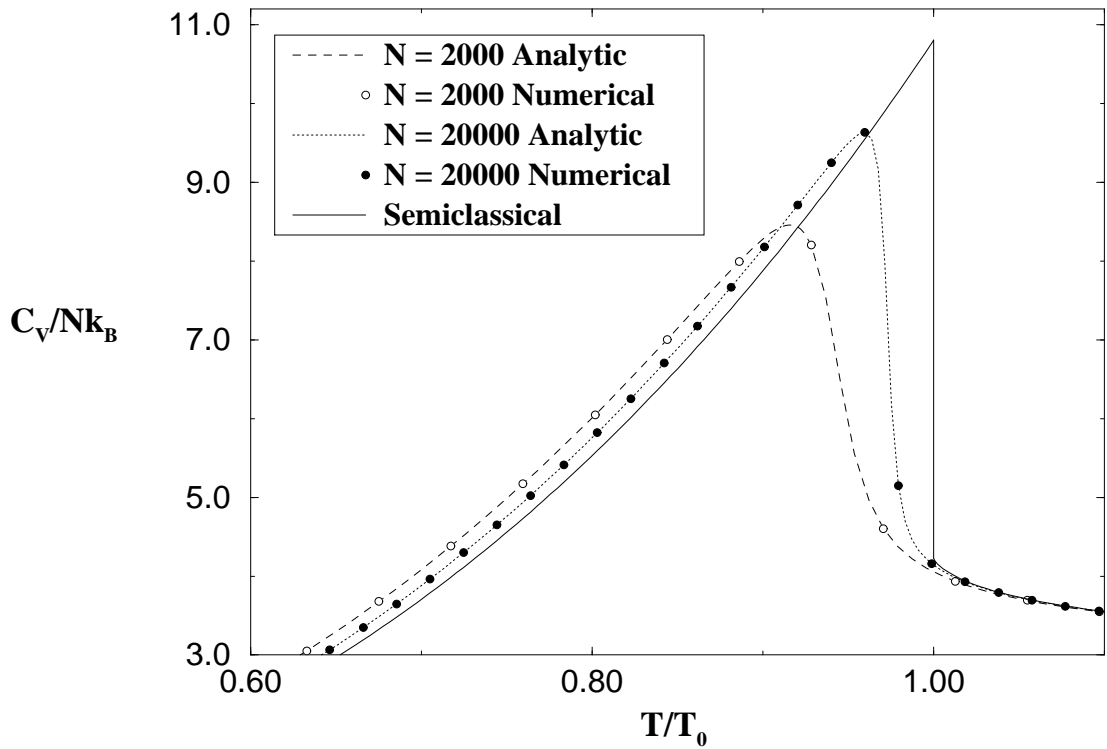,width=14cm,angle=270}}
\end{center}
\caption[b]{\protect\footnotesize The specific heat $C_V/N$ as a function of $T/T_0$ for $N = 2000$ and $N = 20000$. The discontinuity in the specific heat found with the semiclassical approximation is smoothened by the corrections. The sharpness of the peak is seen to increase with $N$.}
\label{SPEC}                                        
\end{figure}

The specific heat is obtained from the internal energy
\beq
     U = \sum_{n=0}^\infty {g_n\ve_n\over e^{\beta(\ve_n - \mu)} - 1}             \label{U1}
\eeq
In the condensate each particle has zero energy so that the sum really starts at the first
excited level with $n = 1$. Again using the Euler-Maclaurin formula (\ref{EM}) we obtain
\beq
     {U\over\hbar\omega} &=& {3\over b^4}Li_4(\lambda e^{-b}) 
       + {6\over b^3}Li_3(\lambda e^{-b}) + {11\over 2b^2}Li_2(\lambda e^{-b}) 
       + {3\over b}Li_1(\lambda e^{-b})    \nn \\
      &+&   {1\over 4}{\lambda e^{-b}\over 1 - \lambda e^{-b}}
      \left({25\over 6} + {b\over 1 - \lambda e^{-b}} \right)                    \label{U2}
\eeq
The specific heat can now be obtained as a function of the fugacity by taking the partial
derivative with respect to temperature. It will depend on the partial derivative
$(\del\lambda/\del T)$ which can be obtained from (\ref{sum3}). The result is shown in Fig. \ref{SPEC}
for different numbers of particles in the trap together with exact numerical results. At
extremely low temperatures the number of excited particles is exponentially small which will
thus result in a correspondingly suppressed specific heat.

When the number of trapped particles is very large, the first term in (\ref{U2}) gives 
the dominant contribution to the result. As a rough approximation, we can keep only this
and setting $\lam \simeq 1$ just below the transition temperature, we have
\beq
    {U\over\hbar\omega} \simeq 3\left({k_BT\over\hbar\omega}\right)^4\zeta(4)      \label{U3}
\eeq
The specific heat at the transition temperature is thus approximately 
\beq 
     {C_V\over Nk_B} \simeq 12 {\zeta(4)\over\zeta(3)}
\eeq
i.e. almost a factor four larger than for a classical gas in the oscillator potential.

By the same Euler-Maclaurin method we have also investigated the anisotropic oscillator trap
where the frequencies $\omega_x,\omega_y$ and $\omega_z$ 
are all different\cite{HHR}. Instead of the above parameter $b$, we now introduce the three
parameters $b_i = \hbar\omega_i/k_BT$. Corresponding to equation (\ref{approx}), we then find 
for the first two leading terms in the result for the number of excited particles,
\beq
    N_e = {1\over b_xb_yb_z}Li_3(\lambda) + {1\over 2}\left({1\over b_xb_y} + {1\over b_yb_z}
        + {1\over b_xb_z}\right)Li_2(\lambda) + \ldots              \label{xyz}
\eeq
Grossmann and  Holthaus\cite{GH} considered a potential with $\omega_x = 600$ Hz, 
$\omega_y = \sqrt{2}\omega_x$ and $\omega_z = \sqrt{3}\omega_x$. They write the coefficient of 
$Li_2(\lambda)$  as $\gamma(k_BT/\hbar\omega)^2$ where $\omega = 
(\omega_x\omega_y\omega_y)^{1/3}$ and find $\gamma \approx 1.6$ from a numerical summation 
over the energy levels of the anisotropic oscillator. On the other hand, we obtain from
(\ref{xyz})
\beq
     \gamma = \left({3\over 4}\right)^{1/3}\left({1\over\sqrt{2}} + {1\over\sqrt{3}} 
            + {1\over\sqrt{6}}\right)   \simeq  1.538
\eeq
which agrees quite well with their approximate result.

We want to thank Mark Burgess for making reference\cite{DT} known to us.

\vspace{10mm}

\bigskip

\end{document}